\pdfoutput=1
\documentclass[aps,prd,showpacs,superscriptaddress,twocolumn,nofootinbib,
]{revtex4-1}

\usepackage{graphicx}  
\usepackage{dcolumn}   
\usepackage{bm}        
\usepackage{amssymb}   
\usepackage{amsmath} 
\usepackage{accents}
\usepackage{physics}
\usepackage{caption}
\usepackage{subcaption}
\usepackage{hyperref}
\usepackage{tabularx,booktabs}
\usepackage{multirow}
\usepackage{xcolor}
\newcolumntype{Y}{>{\centering\arraybackslash}X}
\def\bracketbar{\hbox{\kern-8pt\raise1pt%
  \hbox{{\tiny(}{\lower1.4pt\hbox{\bf--}}{\tiny)}}}}
\begin{document}


\title{Search for light sterile neutrinos with the T2K far detector Super-Kamiokande \\at a baseline of 295\,km}



\newcommand{\INSTHD}{\affiliation{University Autonoma Madrid, Department of Theoretical Physics, 28049 Madrid, Spain}}
\newcommand{\INSTEE}{\affiliation{University of Bern, Albert Einstein Center for Fundamental Physics, Laboratory for High Energy Physics (LHEP), Bern, Switzerland}}
\newcommand{\INSTFE}{\affiliation{Boston University, Department of Physics, Boston, Massachusetts, U.S.A.}}
\newcommand{\INSTD}{\affiliation{University of British Columbia, Department of Physics and Astronomy, Vancouver, British Columbia, Canada}}
\newcommand{\INSTGA}{\affiliation{University of California, Irvine, Department of Physics and Astronomy, Irvine, California, U.S.A.}}
\newcommand{\INSTI}{\affiliation{IRFU, CEA Saclay, Gif-sur-Yvette, France}}
\newcommand{\INSTGB}{\affiliation{University of Colorado at Boulder, Department of Physics, Boulder, Colorado, U.S.A.}}
\newcommand{\INSTFG}{\affiliation{Colorado State University, Department of Physics, Fort Collins, Colorado, U.S.A.}}
\newcommand{\INSTFH}{\affiliation{Duke University, Department of Physics, Durham, North Carolina, U.S.A.}}
\newcommand{\INSTBA}{\affiliation{Ecole Polytechnique, IN2P3-CNRS, Laboratoire Leprince-Ringuet, Palaiseau, France }}
\newcommand{\INSTEF}{\affiliation{ETH Zurich, Institute for Particle Physics, Zurich, Switzerland}}
\newcommand{\INSTEG}{\affiliation{University of Geneva, Section de Physique, DPNC, Geneva, Switzerland}}
\newcommand{\INSTHJ}{\affiliation{University of Glasgow, School of Physics and Astronomy, Glasgow, United Kingdom}}
\newcommand{\INSTDG}{\affiliation{H. Niewodniczanski Institute of Nuclear Physics PAN, Cracow, Poland}}
\newcommand{\INSTCB}{\affiliation{High Energy Accelerator Research Organization (KEK), Tsukuba, Ibaraki, Japan}}
\newcommand{\INSTIB}{\affiliation{University of Houston, Department of Physics, Houston, Texas, U.S.A.}}
\newcommand{\INSTED}{\affiliation{Institut de Fisica d'Altes Energies (IFAE), The Barcelona Institute of Science and Technology, Campus UAB, Bellaterra (Barcelona) Spain}}
\newcommand{\INSTEC}{\affiliation{IFIC (CSIC \& University of Valencia), Valencia, Spain}}
\newcommand{\INSTHH}{\affiliation{Institute For Interdisciplinary Research in Science and Education (IFIRSE), ICISE, Quy Nhon, Vietnam}}
\newcommand{\INSTEI}{\affiliation{Imperial College London, Department of Physics, London, United Kingdom}}
\newcommand{\INSTGF}{\affiliation{INFN Sezione di Bari and Universit\`a e Politecnico di Bari, Dipartimento Interuniversitario di Fisica, Bari, Italy}}
\newcommand{\INSTBE}{\affiliation{INFN Sezione di Napoli and Universit\`a di Napoli, Dipartimento di Fisica, Napoli, Italy}}
\newcommand{\INSTBF}{\affiliation{INFN Sezione di Padova and Universit\`a di Padova, Dipartimento di Fisica, Padova, Italy}}
\newcommand{\INSTBD}{\affiliation{INFN Sezione di Roma and Universit\`a di Roma ``La Sapienza'', Roma, Italy}}
\newcommand{\INSTEB}{\affiliation{Institute for Nuclear Research of the Russian Academy of Sciences, Moscow, Russia}}
\newcommand{\INSTHI}{\affiliation{Institute of Physics (IOP), Vietnam Academy of Science and Technology (VAST), Hanoi, Vietnam}}
\newcommand{\INSTHA}{\affiliation{Kavli Institute for the Physics and Mathematics of the Universe (WPI), The University of Tokyo Institutes for Advanced Study, University of Tokyo, Kashiwa, Chiba, Japan}}
\newcommand{\INSTCC}{\affiliation{Kobe University, Kobe, Japan}}
\newcommand{\INSTCD}{\affiliation{Kyoto University, Department of Physics, Kyoto, Japan}}
\newcommand{\INSTEJ}{\affiliation{Lancaster University, Physics Department, Lancaster, United Kingdom}}
\newcommand{\INSTFC}{\affiliation{University of Liverpool, Department of Physics, Liverpool, United Kingdom}}
\newcommand{\INSTFI}{\affiliation{Louisiana State University, Department of Physics and Astronomy, Baton Rouge, Louisiana, U.S.A.}}
\newcommand{\INSTHB}{\affiliation{Michigan State University, Department of Physics and Astronomy,  East Lansing, Michigan, U.S.A.}}
\newcommand{\INSTCE}{\affiliation{Miyagi University of Education, Department of Physics, Sendai, Japan}}
\newcommand{\INSTDF}{\affiliation{National Centre for Nuclear Research, Warsaw, Poland}}
\newcommand{\INSTFJ}{\affiliation{State University of New York at Stony Brook, Department of Physics and Astronomy, Stony Brook, New York, U.S.A.}}
\newcommand{\INSTGJ}{\affiliation{Okayama University, Department of Physics, Okayama, Japan}}
\newcommand{\INSTCF}{\affiliation{Osaka City University, Department of Physics, Osaka, Japan}}
\newcommand{\INSTGG}{\affiliation{Oxford University, Department of Physics, Oxford, United Kingdom}}
\newcommand{\INSTGC}{\affiliation{University of Pittsburgh, Department of Physics and Astronomy, Pittsburgh, Pennsylvania, U.S.A.}}
\newcommand{\INSTFA}{\affiliation{Queen Mary University of London, School of Physics and Astronomy, London, United Kingdom}}
\newcommand{\INSTE}{\affiliation{University of Regina, Department of Physics, Regina, Saskatchewan, Canada}}
\newcommand{\INSTGD}{\affiliation{University of Rochester, Department of Physics and Astronomy, Rochester, New York, U.S.A.}}
\newcommand{\INSTHC}{\affiliation{Royal Holloway University of London, Department of Physics, Egham, Surrey, United Kingdom}}
\newcommand{\INSTBC}{\affiliation{RWTH Aachen University, III. Physikalisches Institut, Aachen, Germany}}
\newcommand{\INSTFB}{\affiliation{University of Sheffield, Department of Physics and Astronomy, Sheffield, United Kingdom}}
\newcommand{\INSTDI}{\affiliation{University of Silesia, Institute of Physics, Katowice, Poland}}
\newcommand{\INSTIA}{\affiliation{SLAC National Accelerator Laboratory, Stanford University, Menlo Park, California, USA}}
\newcommand{\INSTBB}{\affiliation{Sorbonne Universit\'e, Universit\'e Paris Diderot, CNRS/IN2P3, Laboratoire de Physique Nucl\'eaire et de Hautes Energies (LPNHE), Paris, France}}
\newcommand{\INSTEH}{\affiliation{STFC, Rutherford Appleton Laboratory, Harwell Oxford,  and  Daresbury Laboratory, Warrington, United Kingdom}}
\newcommand{\INSTCH}{\affiliation{University of Tokyo, Department of Physics, Tokyo, Japan}}
\newcommand{\INSTBJ}{\affiliation{University of Tokyo, Institute for Cosmic Ray Research, Kamioka Observatory, Kamioka, Japan}}
\newcommand{\INSTCG}{\affiliation{University of Tokyo, Institute for Cosmic Ray Research, Research Center for Cosmic Neutrinos, Kashiwa, Japan}}
\newcommand{\INSTHF}{\affiliation{Tokyo Institute of Technology, Department of Physics, Tokyo, Japan}}
\newcommand{\INSTGI}{\affiliation{Tokyo Metropolitan University, Department of Physics, Tokyo, Japan}}
\newcommand{\INSTHG}{\affiliation{Tokyo University of Science, Faculty of Science and Technology, Department of Physics, Noda, Chiba, Japan}}
\newcommand{\INSTF}{\affiliation{University of Toronto, Department of Physics, Toronto, Ontario, Canada}}
\newcommand{\INSTB}{\affiliation{TRIUMF, Vancouver, British Columbia, Canada}}
\newcommand{\INSTG}{\affiliation{University of Victoria, Department of Physics and Astronomy, Victoria, British Columbia, Canada}}
\newcommand{\INSTDJ}{\affiliation{University of Warsaw, Faculty of Physics, Warsaw, Poland}}
\newcommand{\INSTDH}{\affiliation{Warsaw University of Technology, Institute of Radioelectronics, Warsaw, Poland}}
\newcommand{\INSTFD}{\affiliation{University of Warwick, Department of Physics, Coventry, United Kingdom}}
\newcommand{\INSTGH}{\affiliation{University of Winnipeg, Department of Physics, Winnipeg, Manitoba, Canada}}
\newcommand{\INSTEA}{\affiliation{Wroclaw University, Faculty of Physics and Astronomy, Wroclaw, Poland}}
\newcommand{\INSTHE}{\affiliation{Yokohama National University, Faculty of Engineering, Yokohama, Japan}}
\newcommand{\INSTH}{\affiliation{York University, Department of Physics and Astronomy, Toronto, Ontario, Canada}}

\INSTHD
\INSTEE
\INSTFE
\INSTD
\INSTGA
\INSTI
\INSTGB
\INSTFG
\INSTFH
\INSTBA
\INSTEF
\INSTEG
\INSTHJ
\INSTDG
\INSTCB
\INSTIB
\INSTED
\INSTEC
\INSTHH
\INSTEI
\INSTGF
\INSTBE
\INSTBF
\INSTBD
\INSTEB
\INSTHI
\INSTHA
\INSTCC
\INSTCD
\INSTEJ
\INSTFC
\INSTFI
\INSTHB
\INSTCE
\INSTDF
\INSTFJ
\INSTGJ
\INSTCF
\INSTGG
\INSTGC
\INSTFA
\INSTE
\INSTGD
\INSTHC
\INSTBC
\INSTFB
\INSTDI
\INSTIA
\INSTBB
\INSTEH
\INSTCH
\INSTBJ
\INSTCG
\INSTHF
\INSTGI
\INSTHG
\INSTF
\INSTB
\INSTG
\INSTDJ
\INSTDH
\INSTFD
\INSTGH
\INSTEA
\INSTHE
\INSTH

\author{K.\,Abe}\INSTBJ
\author{R.\,Akutsu}\INSTCG
\author{A.\,Ali}\INSTBF
\author{C.\,Andreopoulos}\INSTEH\INSTFC
\author{L.\,Anthony}\INSTFC
\author{M.\,Antonova}\INSTEC
\author{S.\,Aoki}\INSTCC
\author{A.\,Ariga}\INSTEE
\author{Y.\,Ashida}\INSTCD
\author{Y.\,Awataguchi}\INSTGI
% ...

\collaboration{The T2K Collaboration}\noaffiliation 


\date{\today}

\begin{abstract}
We perform a search for light sterile neutrinos using the data from the T2K far detector at a baseline of 295\,km, with an exposure of 14.7 (7.6)$\times 10^{20}$ protons on target in neutrino (antineutrino) mode. 
A selection of neutral current interaction samples are also used to enhance the sensitivity to sterile mixing.
No  evidence of  sterile  neutrino mixing  in  the  3+1 model was found from a simultaneous fit to the charged-current muon, electron and neutral current neutrino samples.
We set the most stringent limit on the sterile oscillation amplitude $\sin^2\theta_{24}$ for the sterile neutrino mass splitting $\Delta m^2_{41}<3\times 10^{-3}$\,eV$^2/c^4$.
\end{abstract}


\maketitle

\section{Introduction}
Over the last few decades, the theory of neutrino oscillations has been well established through a series of experiments with neutrinos produced by the Sun\,\cite{Cleveland:1998nv,PhysRevC.80.015807,Altmann:2005ix,Hampel:1998xg,PhysRevC.72.055502,PhysRevD.83.052010}, nuclear reactors\,\cite{PhysRevLett.94.081801,PhysRevLett.108.171803,Abe:2013sxa,Ahn:2012nd}, accelerators\,\cite{PhysRevD.74.072003,Adamson:2013whj,PhysRevLett.112.181801,PhysRevD.89.051102,PhysRevD.93.051104} and in the atmosphere\,\cite{PhysRevLett.81.1562,PhysRevD.91.072004}. Most data from these studies are consistent with the three flavor paradigm where the three weakly interacting neutrino flavors are related to three neutrino mass states by the PMNS mixing matrix\,\cite{Pontecorvo:1967fh,GRIBOV1969493,doi:10.1143}. 
However, deviations from the three flavor scheme have been reported. At LSND\,\cite{PhysRevD.64.112007} and MiniBooNE\,\cite{PhysRevLett.110.161801}, there were excesses of $\bar{\nu}_e$ found in short-baseline $\bar{\nu}_\mu$ beams; MiniBooNE also reported an excess in $\nu_e$ appearance\,\cite{Aguilar-Arevalo:2018gpe}; radioactive calibration sources in gallium experiments\,\cite{Bahcall:1994bq,PhysRevC.73.045805} showed a deficit of $\nu_e$ flux; and reactor experiments\,\cite{PhysRevD.83.073006} observed less $\bar{\nu}_e$ than expected. These results could be explained by a fourth neutrino state with a mass difference $\Delta m^2\sim1$\,eV$^2/c^4$ with respect to the three PMNS states\,\cite{Bilenky:1998dt,Barger:2000ch,Kopp:2013vaa,0954-3899-43-3-033001}. From the measurements of the invisible decay width of the Z$^0$ boson at the LEP collider, the number of weakly interacting neutrino species with mass below 45\,GeV$/c^2$ is limited to three\,\cite{ALEPH:2005ab}, so the new neutrino state must not couple to the weak interaction and is often referred to as $sterile$. We can incorporate this additional neutrino state in the simple ``3+1" model\,\cite{Barger:2000ch}, which involves the three active neutrinos and one sterile neutrino, and study its effect on the oscillation signatures. Currently, the null results, especially in the $\nu^{\bracketbar}_\mu$ disappearance channels, from short-baseline accelerator experiments like CCFR\,\cite{PhysRevLett.52.1384}, MiniBooNE/SciBooNE\,\cite{Cheng:2012yy} and T2K\,\cite{Abe:2014nuo}; long-baseline experiments like MINOS/MINOS+\,\cite{PhysRevLett.122.091803} and NOvA\,\cite{Adamson:2017zcg}; or atmospheric experiments like Super-Kamiokande\,\cite{Abe:2014gda} and IceCube\,\cite{TheIceCube:2016oqi,Aartsen:2017bap}, have limited the available parameter space in the ``3+1" model.

\par The Tokai to Kamioka (T2K) experiment is a long-baseline accelerator neutrino experiment in Japan which primarily measures muon neutrino disappearance and electron neutrino appearance. While T2K is designed for studying standard three flavor oscillation at ${\Delta m^2\sim10^{-3}}$\,eV$^2/c^4$, it also has the potential to search for oscillation signatures due to sterile neutrinos around this ${\Delta m^2}$ range. 
Neutral current (NC) neutrino interactions are also collected  in  the  far detector,  Super-Kamiokande (SK),  which  can  be  used  to 
enhance the sensitivity to sterile mixing as the sterile neutrinos, unlike other active neutrinos, do not interact through CC or NC scattering.
We present a long-baseline search for sterile neutrinos in the ``3+1" framework, using both the charged-current (CC) $\nu^{\bracketbar}_\mu$ and $\nu^{\bracketbar}_e$ samples and NC samples at the far detector. 

\par Sec.\,\ref{sec:mixing} briefly describes the sterile neutrino mixing model and its effect on the oscillation probability
(or oscillation signatures). The T2K experimental setup is outlined in Sec.\,\ref{sec:t2k}, followed by event selection criteria in Sec.\,\ref{sec:evt_sel}. Sec.\,\ref{sec:analysis} explains the analysis strategy and Sec.\,\ref{sec:results} presents our search results. Finally, Sec.\,\ref{sec:conclusions} gives a summary and outlook of our sterile neutrino study.

\section{``3+1" sterile neutrino mixing}\label{sec:mixing}
In this study, we focus on a ``3+1" like model where a single sterile neutrino is added and mixed with the three active states, which is the simplest model with sterile neutrino frequently used in neutrino oscillation analysis.
In this model, there is a new flavor state $\nu_s$ and a new mass state~$\nu_4$ with mass $m_4$ added to the three flavor framework. The relation between the flavor and mass states is given by
\begin{equation}
\ket{\nu_\alpha}=\sum U^*_{\alpha k}\ket{\nu_k},
\end{equation}
where $\ket{\nu_\alpha}$ are the flavor states and $\ket{\nu_k}$ are the mass states. The original $3\times 3$ PMNS mixing matrix is expanded to a $4\times 4$ matrix as:
\begin{equation}
\textbf{U}=\begin{pmatrix}
U_{e1} & U_{e2} & U_{e3} & U_{e4}\\
U_{\mu 1} & U_{\mu 2} & U_{\mu 3} & U_{\mu 4}\\
U_{\tau 1} & U_{\tau 2} & U_{\tau 3} & U_{\tau 4}\\
U_{s1} & U_{s2} & U_{s3} & U_{s4}\\
\end{pmatrix}.
\end{equation}
We choose the parameterization as in\,\cite{HARARI1986123}:
\begin{equation}
\textbf{U}=\textbf{U}_{34}\textbf{U}_{24}\textbf{U}_{14}\textbf{U}_{23}\textbf{U}_{13}\textbf{U}_{12},
\end{equation}
where $\textbf{U}_{ij}$ is a unitary rotation matrix of an angle $\theta_{ij}$ in the $ij$-plane. There are therefore three new mixing angles $\theta_{14}$, $\theta_{24}$, $\theta_{34}$ and two new CP-violating phases $\delta_{14}$, $\delta_{24}$. Note that $\sin^2\theta_{14}$ has been constrained to small values by reactor experiments\,\cite{PhysRevLett.117.151801}, and T2K has limited sensitivity to $\theta_{14}$ and the new CP phases. Since there is no significant correlation between them and the other oscillation parameters in this study, we set $\theta_{14}=\delta_{14}=\delta_{24}=0$ to simplify the mixing matrix.
\par
\begin{figure}
	\centering
	\includegraphics[width=0.9\linewidth]{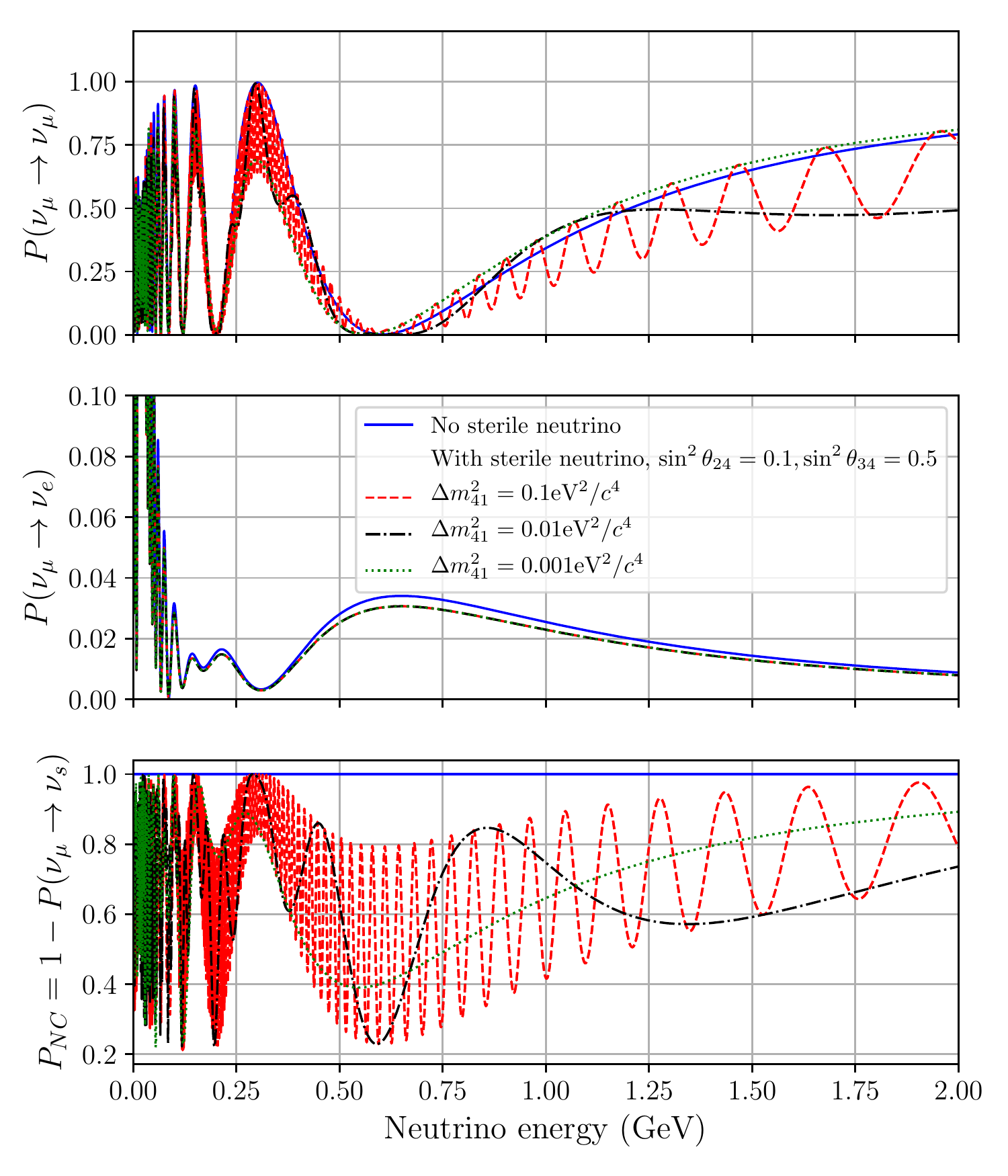}
	\caption{The muon neutrino survival probability (top), electron neutrino appearance probability (middle) and active neutrino survival probability (bottom) as a function of neutrino energy, with and without sterile neutrino, assuming $\Delta m^2_{31}>0$.}\label{fig:p_osc}
\end{figure}

At the far detector, 
the $\nu_\mu$ survival probability can be approximated (omitting $\delta_{CP}$ terms) as:
\begin{equation}\label{eq:pmumu}
\begin{split}
P(\nu_\mu\rightarrow\nu_\mu)\approx  1 &-\sin^22\theta_{23}\cos^4\theta_{24}\sin^2\frac{\Delta m^2_{31}L}{4E}
\\ & - \cos^2\theta_{23}\sin^22\theta_{24}\sin^2\frac{\Delta m^2_{41}L}{4E},
\\ & - \sin^2\theta_{23}\sin^22\theta_{24}\sin^2\frac{\Delta m^2_{43}L}{4E},
\end{split}
\end{equation}
and the $\nu_e$ appearance probability as:
\begin{equation}\label{eq:pmue}
P(\nu_\mu\rightarrow\nu_e)\approx \sin^22\theta_{13}\cos^2\theta_{24}\sin^2\theta_{23}\sin^2\frac{\Delta m^2_{31}L}{4E}.
\end{equation}
Thus the CC channels are sensitive to $\theta_{24}$ and $\Delta m^2_{41}$. Similarly, the active neutrino survival probability, which is manifested in the NC channel, is sensitive to $\theta_{24}$, $\Delta m^2_{41}$, and $\theta_{34}$:
\begin{equation}\label{eq:pmus}
\begin{split}
P_{NC}= 1 &- P(\nu_\mu\rightarrow\nu_s)
\\\approx 1 &- \sin^22\theta_{23}(A^2-\frac{1}{4}B^2)\sin^2\frac{\Delta m^2_{31}L}{4E}
\\          &- B(B\cos^2\theta_{23}-A\sin2\theta_{23})\sin^2\frac{\Delta m^2_{41}L}{4E}
\\          &- B(B\sin^2\theta_{23}+A\sin2\theta_{23})\sin^2\frac{\Delta m^2_{43}L}{4E},
\end{split}
\end{equation}
where ${A=\cos\theta_{24}\sin\theta_{34}}$ and ${B=\sin 2\theta_{24}\cos\theta_{34}}$. The antineutrino oscillation probabilities follow similarly, but there are small differences due to the $\delta_{CP}$ terms which are not explicitly written here.
Figure\,\ref{fig:p_osc} shows schematically how the oscillation probabilities are modified with the mixing of sterile neutrinos.

The addition of a sterile neutrino state which experiences no matter potential (both CC and NC) complicates the calculation of oscillation in matter. We estimated the possible impacts due to matter effects at T2K baseline and energy with numerical calculation of oscillation probabilities using nuCraft\,\cite{Wallraff:2014qka}, and found at most a few percent changes on the $\nu^{\bracketbar}_e$ and NC samples. This is negligible at current precision, so we simply use the vacuum oscillation probabilities (without approximation) in this study.

\section{The T2K experiment}\label{sec:t2k}
The T2K experiment\,\cite{ABE2011106} consists of a neutrino beam, a near detector complex, and the water Cherenkov detector Super-Kamiokande\,\cite{FUKUDA2003418} as the far detector at a baseline of 295\,km. T2K is sensitive to  $\theta_{13}$ and $\delta_{CP}$ through the $\nu^{\bracketbar}_e$ appearance searches, and can also make precision measurements on $\theta_{23}$ and mass difference $\Delta m^2_{32}$ with the $\nu^{\bracketbar}_\mu$ disappearance samples.

\par
The main ring accelerator in  the  Japan Proton  Accelerator Research Complex (J-PARC) produces a 30\,GeV proton beam with spills every 2.48\,s that contain eight beam bunches which are 580\,ns apart. The proton 
beam is extracted to the neutrino beamline and collides with a graphite target to produce charged pions and kaons.
In the neutrino (antineutrino) beam mode, magnetic horns are used to focus the positively (negatively)-charged pions and kaons into a 96\,m long decay volume filled with helium gas. These mesons typically decay into muon neutrinos (antineutrinos). The neutrino beamline is directed at an angle of 2.5$^\text{o}$ away from the far detector, so that the off-axis beam at SK has a narrower peak at 0.6\,GeV with much less $\nu^{\bracketbar}_e$ content than an on-axis beam.

\par
The near detector complex located at 280\,m from the neutrino production target contains two detectors. The on-axis detector, INGRID, is an array of 16 iron/scintillator detectors which precisely measure the beam direction and intensity\,\cite{Abe:2011xv}. The off-axis detector, ND280, is a magnetized tracking detector which constrains the neutrino flux and cross-section model parameters in the oscillation analysis. 

\par
The far detector, Super-Kamiokande, is located in Gifu prefecture, at a distance of 295 km from the J-PARC neutrino 
beam. It is a water Cherenkov detector consisting of 50\,kt of ultra-pure water. 
The tank is optically separated into two regions. The inner detector (ID) is a cylinder containing 32\,kt of water and is instrumented with 11,129 inward-facing 20\,inch photomultiplier tubes (PMTs). The outer detector (OD) extends 2\,m outward from the ID and is instrumented with 1885 outward-facing 8\,inch PMTs. The OD serves as an active veto against cosmic-ray muons, and provides passive shielding from radioactivity in the surrounding rock. 
The expected event rates at far detector are calculated by multiplying the unoscillated neutrino spectra (predicted by near detectors) with the corresponding oscillation probabilities.


\section{Event selection at far detector}\label{sec:evt_sel}
The T2K Runs 1--8 data set used in this analysis was collected from Jan. 2010 to May 2017, corresponding to a beam exposure of $14.7\times 10^{20}$ protons on target (POT) in neutrino mode and $7.6\times 10^{20}$\,POT in antineutrino mode.
Events at the far detector are required to occur within 1\,ms of the beam spill time window, and be fully contained in the fiducial volume of the SK ID. 
For the CC $\nu^{\bracketbar}_\mu$ and $\nu^{\bracketbar}_e$ samples, a new Cherenkov-ring reconstruction algorithm\,\cite{PhysRevD.91.072010} is used to select neutrino events, which improves signal/background discrimination and expands the fiducial volume to increase statistics. 

\begin{table}
	\caption{\label{tab:evtRate}%
		Number of events expected and observed in the eight oscillation samples used in this analysis. Three flavor oscillation is assumed in expected rate.
	}
	\begin{ruledtabular}
		\begin{tabular}{lcc}
			Sample & Expected & Observed\\
			\colrule
			$\nu_\mu$ CC-0$\pi$  & 268.4 & 240 \\
			$\bar{\nu}_\mu$ CC-0$\pi$  & 64.3 & 68 \\
			$\nu_e$ CC-0$\pi$  & 73.5 & 74 \\
			$\bar{\nu}_e$ CC-0$\pi$  & 7.9 & 7 \\
			$\nu_e$ CC-1$\pi^+$ & 6.9 & 15\\
			\colrule
			$\nu$ NC$\pi^0$ & 49.5 & 53 \\
			$\bar{\nu}$ NC$\pi^0$ & 11.3 & 9\\
			NC $\gamma$-deexcit. & \multirow{ 2}{*}{107.7} & \multirow{ 2}{*}{102}\\
			(Runs 1--4) & &\\
		\end{tabular}
	\end{ruledtabular}
\end{table}	

\par
There are five CC analysis samples that are commonly used in the standard three flavor oscillation analysis\,\cite{Abe:2018wpn}: $\nu^{\bracketbar}_\mu$ CC-0$\pi$ and $\nu^{\bracketbar}_e$ CC-0$\pi$ samples which are enriched in CC quasi-elastic (CCQE) events, and $\nu_e$ CC-1$\pi^+$ sample where a $\pi^+$ below Cherenkov threshold is produced. The $\nu^{\bracketbar}_\mu$ samples are binned in reconstructed neutrino energy, and the $\nu^{\bracketbar}_e$ samples in reconstructed lepton momentum and angle $\theta$ relative to the beam. Details can be found in \cite{Abe:2018wpn}. Table\,\ref{tab:evtRate} summarizes the event rates, where the Monte Carlo (MC) expectation is calculated with $\sin^2\theta_{23}=0.528$, $\Delta m^2_{32}=2.509\times 10^{-3}$\,eV$^2/c^4$, $\delta_{CP}=-1.601$ (the most probable values obtained by the Bayesian analysis in\,\cite{PhysRevD.91.072010}), and $\sin^2\theta_{13}=0.0219$ (taken from\,\cite{1674-1137-40-10-100001}). All sterile mixing angles are set to zero. 

\par
In addition to the CC samples, beam-induced NC events are also collected in SK. These events have previously only been used in publications for systematic uncertainties\,\cite{Abe:2018wpn} and  cross-section\,\cite{Abe:2014dyd} studies.
In  this  analysis,  NC$\pi^0$ and NC $\gamma$-deexcitation samples are used in the oscillation fit to enhance the 
sensitivity to sterile mixing parameters.

\par
The NC$\pi^0$ samples select neutrino events with single $\pi^0$ production, where $\pi^0\rightarrow 2\gamma$ decay produces two visible Cherenkov rings in the detector. 
Events with two electron-like Cherenkov rings are selected as candidates for NC$\pi^0$ samples, and those with decay electron candidates (from muons) are rejected. The invariant mass from the two rings is required to be between 85\,MeV$/c^2$ and 
135\,MeV$/c^2$ to be consistent with the $\pi^0$ mass. 
From simulations, 68.5\% (53.6\%) of events originate from a $\Delta$ resonance and 19.1\% (34.9\%) from coherent pion production for (anti)neutrino mode. 
The NC single pion resonant (NC1$\pi$) production is described by the Rein-Sehgal model\,\cite{REIN198179}, while the coherent production is described by a tuned model of Rein-Sehgal\,\cite{REIN198329}. 
In MC, NC events constitute 97.1\% (98.5\%) of the sample. 

\par
The NC $\gamma$-deexcitation sample was first reported in the measurement of neutrino-oxygen NC quasi-elastic (NCQE) cross-section\,\cite{Abe:2014dyd}. The NCQE cross-section is calculated by a spectral function model\,\cite{Benhar:2005dj,BENHAR1994493} with the BBBA05 form factor parameterization\,\cite{Bradford:2006yz}, reweighting as a function of neutrino energy to match the theoretical calculations\,\cite{Ankowski:2011ei}. The NCQE interaction can knockout a nucleon,
\[
\begin{split}
\nu+^{16}\text{O}&\rightarrow\nu+p+^{15}\text{N}^*, \text{ or}
\\\nu+^{16}\text{O}&\rightarrow\nu+n+^{15}\text{O}^*,
\end{split}
\]
which produces primary $\gamma$ rays from residual nucleus deexcitation and secondary $\gamma$ rays when knocked-out nucleons interact with other nuclei in water. The emitted $\gamma$ rays are 10\,MeV per event on average, which is much less energetic than other samples. Momenta and vertex positions are reconstructed using the low-energy tools developed for SK solar neutrino analyses\,\cite{Abe:2010hy}. Cuts on reconstructed energy, fiducial volume, event timing, vertex and reconstruction quality, detector pre-activity, and Cherenkov opening angle are applied sequentially to remove beam-unrelated (e.g. radioactivity) and beam-related CC backgrounds. The sample is estimated to contain 76.9\% NCQE and 17.6\% NC non-QE events. The NC $\gamma$-deexcitation sample is currently available only for T2K Runs 1--4 (from Jan. 2010 to May 2013), corresponding to $6.56\times 10^{20}$\,POT. Rest of the data is under reduction and validation with an improved event selection process. 

\par
\begin{figure}
	\centering
	\begin{subfigure}{\linewidth}
		\includegraphics[width=\textwidth]{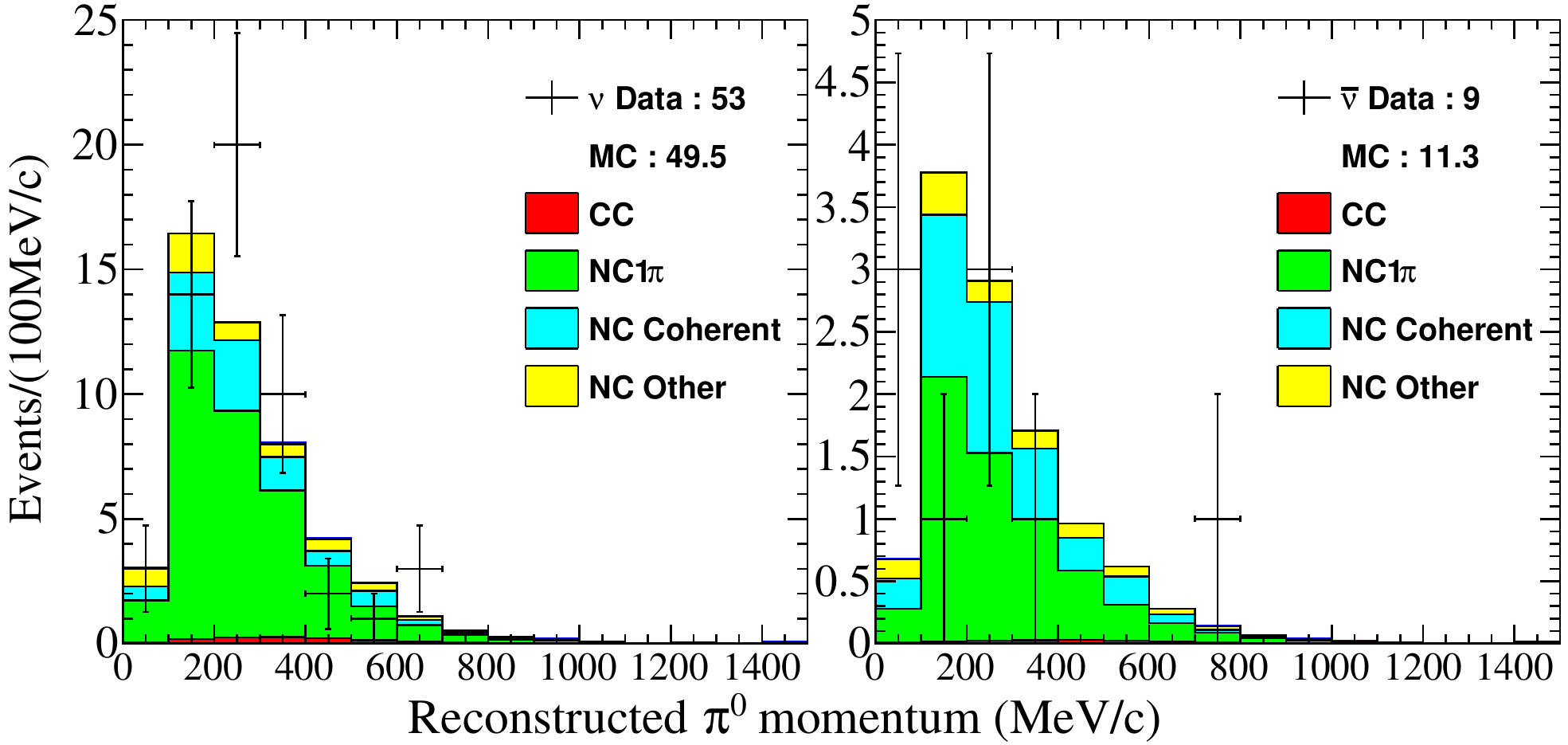}
		\caption{NC$\pi^0$ sample}
	\end{subfigure}
	\begin{subfigure}{0.7\linewidth}
		\includegraphics[width=\textwidth]{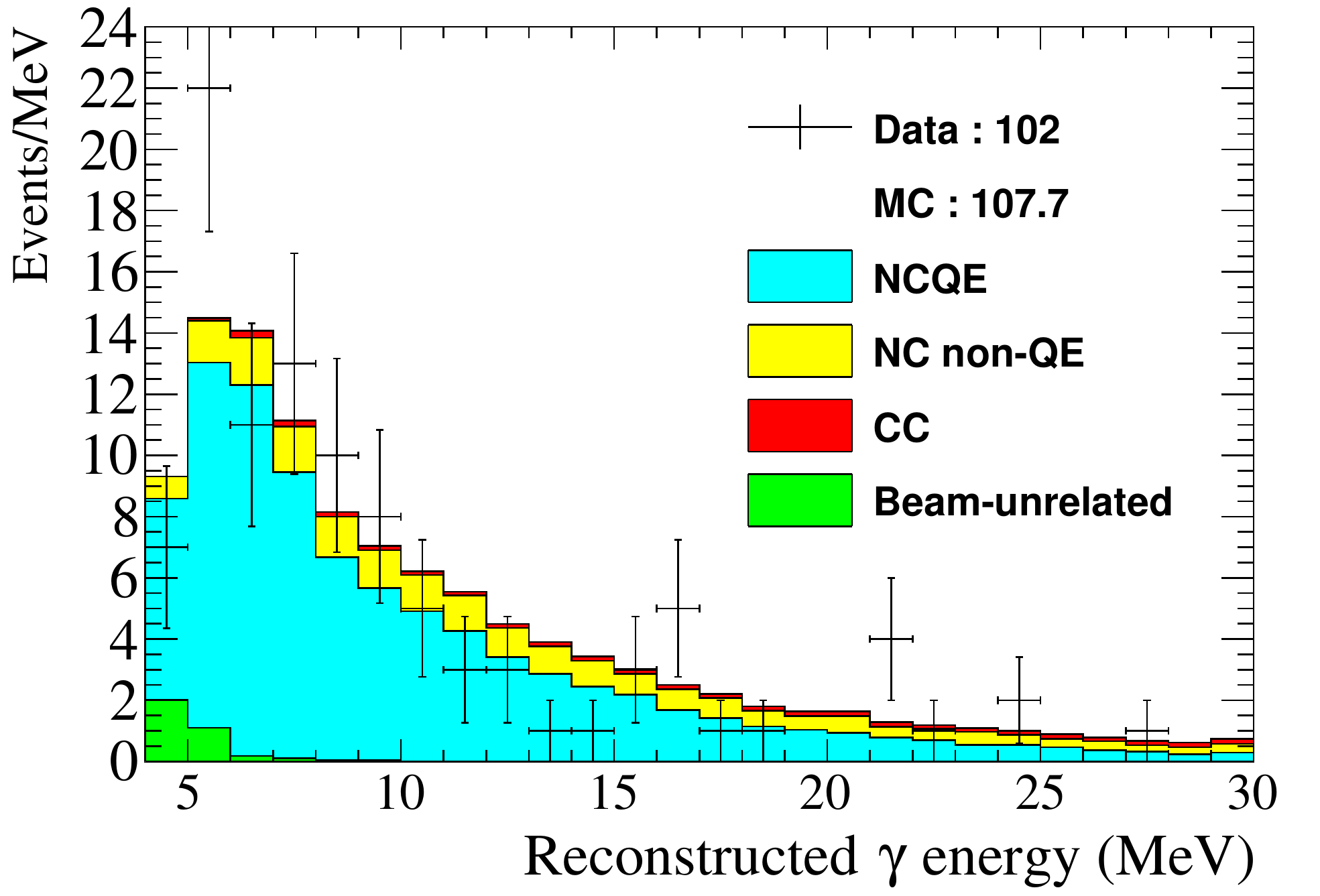}
		\caption{NC $\gamma$-deexcitation sample (Runs 1--4)}
	\end{subfigure}
	\caption{
		Reconstructed $\pi^0$ momentum spectra of NC$\pi^0$ samples (top), and reconstructed gamma energy spectrum of NC $\gamma$-deexcitation sample (bottom). The numbers of events in data and MC prediction are shown in the legend.
	}\label{fig:nc_evt_spec}
\end{figure}

Figure\,\ref{fig:nc_evt_spec} shows the reconstructed $\pi^0$ momentum and $\gamma$ energy distributions of the NC$\pi^0$ and NC $\gamma$-deexcitation samples respectively, with event rates summarized in Table\,\ref{tab:evtRate}. Since the event spectra have little information about true neutrino energy, the sensitivity to $\Delta m^2_{41}$ is limited in the NC channel.

\section{Analysis strategy}\label{sec:analysis}
The overall analysis method is similar to that used in the standard T2K three flavor analysis\,\cite{Abe:2018wpn}, including the incorporation of off-axis near detector data and treatment of systematic uncertainties. However, this analysis performs a simultaneous fit to the five CC and three newly added NC samples to constrain the sterile mixing parameters in the ``3+1" model. Systematic parameters are added to accommodate the possible uncertainties in the NC samples.
\par
The neutrino flux is calculated by a data-driven simulation\,\cite{PhysRevD.96.092006,PhysRevD.87.012001}, which incorporates the conditions of the proton beam, magnetic horn current and neutrino beam-axis direction. Hadronic interactions are tuned with the thin target measurements in the NA61/SHINE experiment\,\cite{Abgrall2016}.  Stability of the neutrino flux has been monitored by INGRID throughout the whole data taking period. At the peak energy 0.6\,GeV, the (anti)neutrino mode beam contains 97.2\% (96.2\%) $\nu^{\bracketbar}_\mu$, with only 0.42\% (0.46\%) $\nu^{\bracketbar}_e$ contamination, and the flux uncertainty is approximately 9\%.
\par
Neutrino events at the near and far detectors are generated by the NEUT 5.3.2 neutrino interaction generator\,\cite{Hayato:2009zz}, which accounts for general interaction and cross-section effects.
Most of the cross-section and neutrino flux parameters are constrained by ND280. The unoscillated CC candidate events at ND280 are classified into different samples according to the event topology, and is fit with a binned Poisson likelihood to extract the best-fit parameters and correlated uncertainties. The central values and their covariances are then propagated to estimate the far detector flux and cross-section parameters and uncertainty covariance matrix. The fit to ND data was done assuming no oscillation at ND280. This approximation is valid for small $\Delta m^2_{41}$, below around 0.3\,eV$^2$/c$^4$. 
However, NC and $\nu_e$ interaction parameters are not constrained by the ND280 fit. 
As a result, an additional uncorrelated 30\% normalization uncertainty is used in this analysis for the NC1$\pi$ and NCQE channels. 
The values of these uncertainties are conservative estimates determined from a previous cross-section analysis\,\cite{Abe:2015awa} and NCQE theoretical model comparisons\,\cite{Ankowski:2015lma,Ankowski:2013aqa}. 
They therefore dominate the overall cross-section uncertainty in the NC oscillation samples. 
\par
At the far detector, there are NEUT parameters (present in ND280 fit but not constrained) that control the final state interactions inside nuclei and secondary interactions with water molecules,
altering the event topologies of outgoing particles. 
For each event topology, the SK selection efficiency and mis-identification rate is separately parameterized. 
The NC $\gamma$-deexcitation sample has separate uncertainties related to primary and secondary $\gamma$ production\,\cite{Abe:2014dyd}. The uncertainties in these parameters constitute the SK detector uncertainties.
\par
The effects of systematic uncertainties on the predicted event rates are summarized in Table\,\ref{tab:sys_evtRate}. The NC cross-section errors are dominant in the NC samples. 
\begin{table}
	\caption{\label{tab:sys_evtRate}%
		Percentage systematic uncertainty on far detector event yields.
	}
	\begin{ruledtabular}
		\begin{tabular}{l|ccc|c}
			Sample & Flux & Cross-sec. & SK Detector & Total\\
			\colrule
			$\nu_\mu$ CC-0$\pi$  & 4.1 & 4.7 & 3.3 & 4.8\\
			$\bar{\nu}_\mu$ CC-0$\pi$  & 3.8 & 4.0 & 2.9 & 4.1\\
			$\nu_e$ CC-0$\pi$  & 4.3 & 5.5 & 3.8 & 6.4\\
			$\bar{\nu}_e$ CC-0$\pi$  & 3.9 & 5.2 & 4.3 & 6.4\\
			$\nu_e$ CC-1$\pi^+$ & 4.3 & 5.0 & 17.1 & 17.7\\
			\colrule
			$\nu$ NC$\pi^0$ & 4.2 & 20.1 & 8.8 & 21.3\\
			$\bar{\nu}$ NC$\pi^0$ & 3.8 & 19.1 & 8.6 & 20.4\\
			NC $\gamma$-deexcit. & 4.1 & 21.1 & 13.2 & 23.3\\
		\end{tabular}
	\end{ruledtabular}
\end{table}
\par
Using the flux and cross-section inputs from ND280, the unoscillated event sample spectra at the far detector are calculated. Oscillation parameters are varied to obtain the best agreement between data and predicted event rates. A joint maximum-likelihood fit to eight far detector samples constrains the sterile mixing parameters $\sin^2\theta_{24}$, $\sin^2\theta_{34}$ and  $\Delta m^2_{41}$. 
The log-likelihood is defined as
\begin{equation}\label{eq:likelihood}
-\ln\mathcal{L}=\sum_{i}[\mu_i-n_i+n_i\ln(n_i/\mu_i)]+\frac{1}{2}\Delta\vec{f}^TV^{-1}\Delta\vec{f},
\end{equation}
where $n_i$ is the number of events in the $i$-th data bin, and $\mu_i=\mu_i(\vec{\theta},\vec{f})$ is the expected event rate with oscillation parameters $\vec{\theta}$ and systematic parameters $\vec{f}$. The last term in Eq.\,\ref{eq:likelihood} accounts for the systematic penalty with $\Delta\vec{f}$ being the difference between the systematic parameters and their prior values, related by the covariance matrix V. 
The oscillation parameters $\sin^2\theta_{23}$, $|\Delta m^2_{32}|$ and $\delta_{CP}$ are allowed to vary without constraint; $\theta_{12}$ and $\Delta m^2_{21}$ are fixed to their PDG values\,\cite{1674-1137-40-10-100001}; and a penalty term is used to constrain $\sin^22\theta_{13}=0.0857\pm 0.0046$.
\par
During the fitting process, at a grid point of $(\sin^2\theta_{24},\sin^2\theta_{34},\Delta m^2_{41})$, the function in Eq.\,\ref{eq:likelihood} is minimized with respect to the other oscillation parameters and systematic parameters. We use  Wilks's theorem to 
estimate the confidence levels (C.L.)\,\cite{wilks1938}. The results are cross-checked with Gaussian CL$_\text{s}$ contours\,\cite{Qian:2014nha} to ensure no significant bias due to the physical limit of $\sin^2\theta_{24}\geq 0$ and $\sin^2\theta_{34}\geq 0$.

\section{Results}\label{sec:results}
We consider the parameter space of $\Delta m^2_{41}>\Delta m^2_{21}$ which is most sensitive in T2K. Two categories of fits are done for normal (NH, $\Delta m^2_{31}>0$) and inverted (IH, $\Delta m^2_{31}<0$) neutrino mass hierarchies respectively. The case of $\Delta m^2_{41}<0$ is very similar and can be obtained by flipping the hierarchy. The ``3+1" best-fit differs from the standard three flavor best-fit by $\Delta\chi^2=1.0$ (4.7) for NH (IH).
From 2500 sets of MC studies with statistical fluctuations, this level of disagreement is expected with the standard three flavor hypothesis in 50\% (30\%) of the studies.

\par
\begin{figure}
	\centering
	\includegraphics[width=0.75\linewidth]{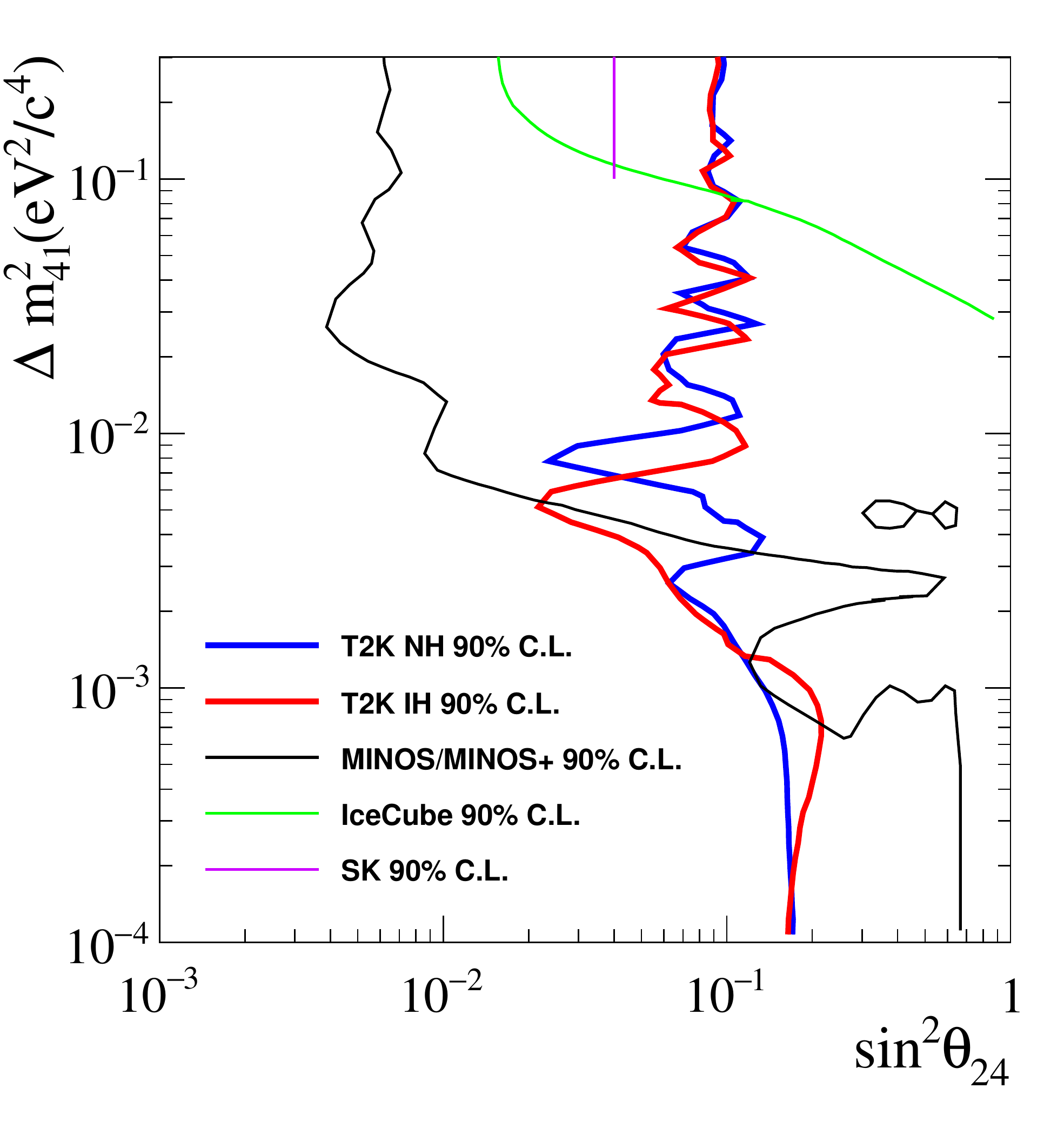}
	\caption{The T2K 90\% exclusion limits on $\sin^2\theta_{24}$ as a function of $\Delta m^2_{41}$, with results from other experiments \cite{PhysRevLett.122.091803,Abe:2014gda,TheIceCube:2016oqi}. The areas on the right are excluded.}\label{fig:sensi_th24_dm41}
\end{figure}

In the $(\sin^2\theta_{24},\Delta m^2_{41})$ parameter plane, $\sin^2\theta_{24}$ is scanned from $10^{-3}$ to 1, and $\Delta m^2_{41}$ from $10^{-4}$ to 0.3\,eV$^2/c^4$. For larger values of $\Delta m^2_{41}$, oscillations would be also seen at the near detectors, which is beyond the scope of this analysis. 
Figure\,\ref{fig:sensi_th24_dm41} shows the T2K 90\% exclusion limits together with results from other experiments. We have set the most stringent limit on $\sin^2\theta_{24}$ for $\Delta m^2_{41}<3\times 10^{-3}$\,eV$^2/c^4$. In particular, the NC samples improve the limit by around 20\% for ${\Delta m^2_{41}< 10^{-3}}$\,eV$^2/c^4$. The limit is weaker at larger $\Delta m^2_{41}$ due to the lack of high energy events, resulting from the sharply peaked off-axis neutrino flux. 
The difference between NH and IH comes from the $\Delta m^2_{43}$ oscillation term. It becomes particularly important when $\Delta m^2_{41}\sim\Delta m^2_{31}$, as this results in very different values of $\Delta m^2_{43}$ in NH and IH.  In partly-degenerate cases where  $\Delta m^2_{41}$ and $\Delta m^2_{31}$ are in integer multiples, the $\nu^{\bracketbar}_e$ and NC samples are important in resolving ambiguities.  In cases where one of the $\Delta m^2$ values is very small, matter effects can significantly alter which mass states are involved in the oscillation, but the overall $\nu^{\bracketbar}_e$ appearance probability is  not affected by more than few percent so this does not significantly modify the exclusion limits.
\par
\begin{figure}
	\centering
	\begin{subfigure}{0.75\linewidth}
		\includegraphics[width=\textwidth]{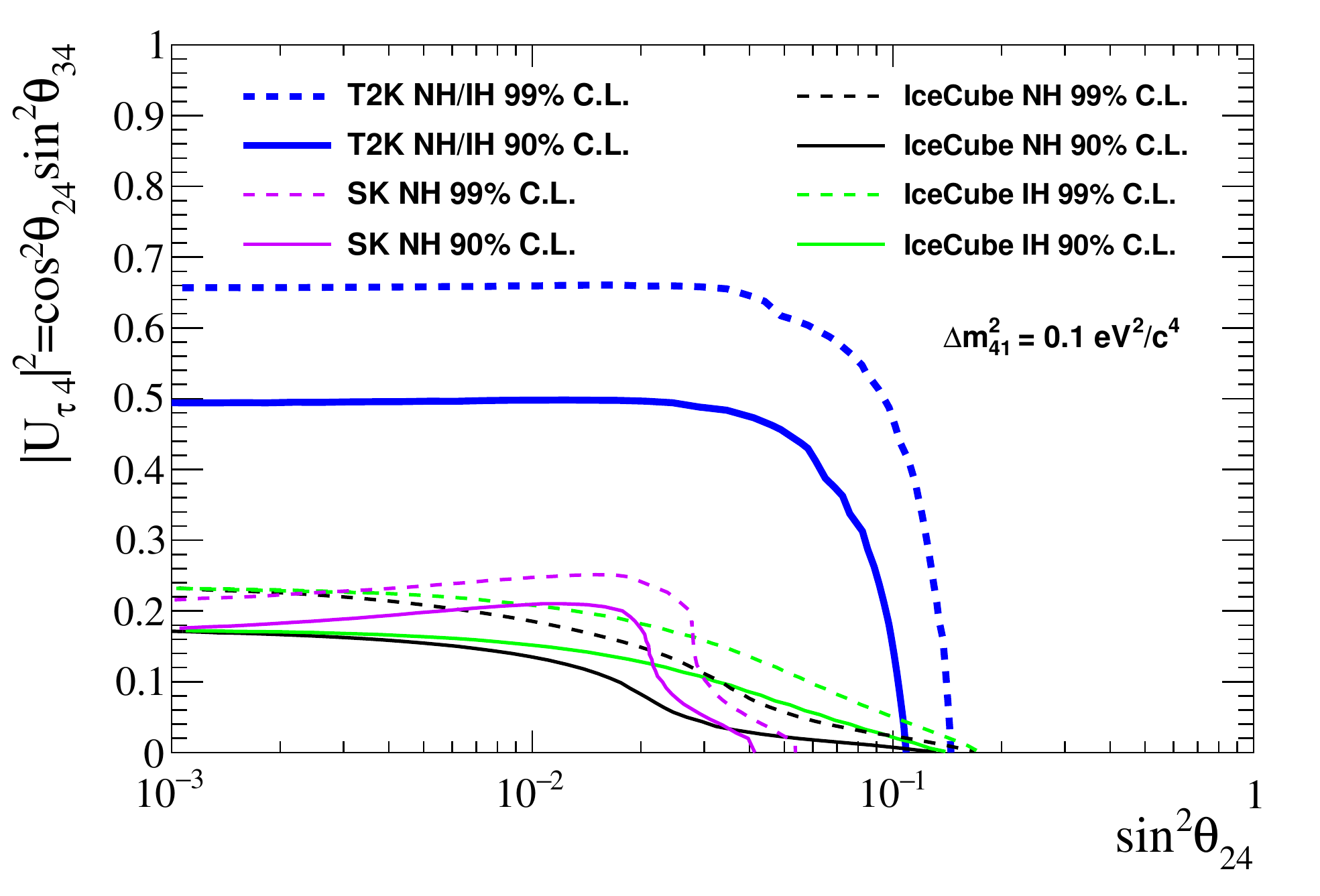}
	\end{subfigure}
	\begin{subfigure}{0.75\linewidth}
		\includegraphics[width=\textwidth]{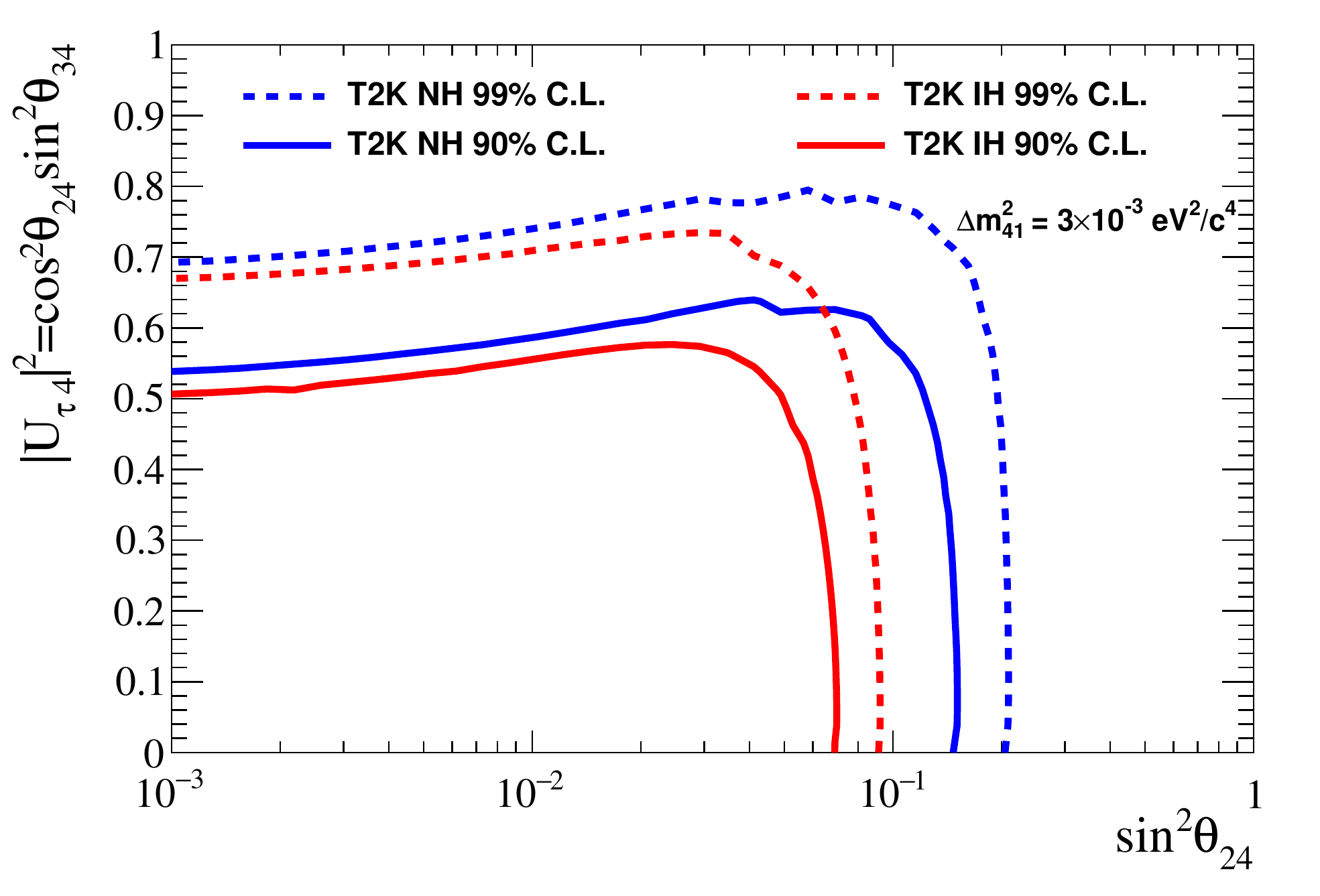}
	\end{subfigure}
	\caption{The T2K 90\% and 99\% exclusion limits on $\sin^2\theta_{24}$ and $|U_{\tau 4}|^2=\cos^2\theta_{24}\sin^2\theta_{34}$ at $\Delta m^2_{41}=0.1$\,eV$^2$ (top), with results from other experiments\,\cite{Aartsen:2017bap,Abe:2014gda}, and exclusion limits at $\Delta m^2_{41}=3\times 10^{-3}$\,eV$^2$ (bottom). The areas on the right are excluded.}\label{fig:sensi_th24_th34}
\end{figure}

The NC samples allow us to constrain $\theta_{34}$ in conjunction with $\theta_{24}$.  Because these samples have low statistics and large cross section uncertainties, we have limited sensitivity, but our results are consistent with other measurements.  Figure\,\ref{fig:sensi_th24_th34} shows that we constrain $\sin^2\theta_{24}<0.1$ and $|U_{\tau 4}|^2=\cos^2\theta_{24}\sin^2\theta_{34}<0.5$ at 90\%~C.L. if  ${\Delta m^2_{41}=0.1}$\,eV$^2/c^4$ is assumed. At smaller $\Delta m^2_{41}$ values, the limits are different between NH and IH.
\section{Conclusions}\label{sec:conclusions}
Data collected by the T2K experiment between 2010 and 2017 (T2K Runs 1--8) have been used to search for oscillation signatures due to light sterile neutrinos in the ``3+1" model. The sterile mixing parameters $(\sin^2\theta_{24},\sin^2\theta_{34}, \Delta m^2_{41})$ are constrained by performing a joint fit of the five CC samples: $\nu^{\bracketbar}_\mu$ CC-0$\pi$, $\nu^{\bracketbar}_e$ CC-0$\pi$, $\nu_e$ CC-1$\pi^+$, and the three new NC samples: $\nu^{\bracketbar}$ NC$\pi^0$, NC $\gamma$-deexcitation, selected at the far detector. 
Systematic uncertainties on the neutrino flux and CC interaction cross-section are constrained by the ND280 data, while NC cross-section uncertainties are determined from a comparison of theoretical models and external data. 
The data are consistent with the standard three flavor oscillation hypothesis. Limits have been set on the sterile mixing parameters, with the world's best constraint on $\sin^2\theta_{24}$ for $10^{-4}$\,eV${^2/c^4<\Delta m^2_{41}<3\times 10^{-3}}$\,eV$^2/c^4$. The data related to the measurement and results presented in this paper can be found in\,\cite{data}.
\par
Our current precision is restricted by statistics and the uncertainty on the NC interaction cross-section. Apart from future updates of the analysis as we take more data,
dedicated systematic studies are required for further improvements to the precision. 
Another possible extension is to perform a joint analysis of near and far detector data that would expand the range of constraint to ${\Delta m^2_{41}\gtrsim 1}$\,eV$^2/c^4$ with additional data at smaller $L/E$.

\begin{acknowledgments}
We thank the J-PARC staff for superb accelerator performance. We thank the CERN NA61/SHINE Collaboration for providing valuable particle production data. We acknowledge the support of MEXT, Japan; NSERC (Grant No. SAPPJ-2014-00031), NRC and CFI, Canada; CEA and CNRS/IN2P3, France; DFG, Germany; INFN, Italy; National Science Centre (NCN) and Ministry of Science and Higher Education, Poland; RSF, RFBR, and MES, Russia; MINECO and ERDF funds, Spain; SNSF and SERI, Switzerland; STFC, UK; and DOE, USA. We also thank CERN for the UA1/NOMAD magnet, DESY for the HERA-B magnet mover system, NII for SINET4, the WestGrid and SciNet consortia in Compute Canada, and GridPP in the United Kingdom. In addition, participation of individual researchers and institutions has been further supported by funds from ERC (FP7), "la Caixa” Foundation (ID 100010434, fellowship code LCF/BQ/IN17/11620050), the European Union’s Horizon 2020 Research and Innovation programme under the Marie Sklodowska-Curie grant agreement no. 713673 and H2020 Grant No. RISE-GA644294-JENNIFER 2020; JSPS, Japan; Royal Society, UK; the Alfred P. Sloan Foundation and the DOE Early Career program, USA.
\end{acknowledgments} 
 
\bibliography{ms}

\end{document}